\documentclass[10pt,twocolumn]{IEEEtran}
\usepackage{cite}
\usepackage{pslatex,bm}
\usepackage{epsfig}
\usepackage{amsmath}
\usepackage{amssymb}
\usepackage{array}
\usepackage{multicol}
\usepackage{color}
\usepackage{url}
\usepackage{dsfont}
\psfull

\newtheorem{lemma}{Lemma}

\newtheorem{definition}{Definition}
\newtheorem{coro}{Corollary}

\DeclareMathAlphabet{\mathpzc}{OT1}{pzc}{m}{it}

\renewcommand{\baselinestretch}{1}
\def\calW{{\mathcal W}}

\def\ba{{\mathbf a}}
\def\bb{{\mathbf b}}

\def\be{{\mathbf e}}

\def\bI{{\mathbf I}}
\def\bK{{\mathbf K}}

\def\bM{{\mathbf M}}

\def\bT{{\mathbf T}}

\def\tr{{\rm tr}}

\def\b0{{\mathbf 0}}

\begin{document}

% paper title
\title{On the Secrecy Rate Region of a Fading Multiple-Antenna Gaussian Broadcast Channel with Confidential Messages and Partial CSIT}

\author{Pin-Hsun Lin\hspace{1.8cm} Chien-Li Su\hspace{1.8cm} Hsuan-Jung Su\\

\normalsize{\it pinhsunlin@gmail.com} \hspace{0.3cm} {\it theone21cent@gmail.com\mbox{    }     }
\vspace{0.5cm}{\it hjsu@cc.ee.ntu.edu.tw}\\
Graduate Institute of Communications Engineering, Department of Electrical Engineering, \\
National Taiwan University, Taipei, Taiwan 10617\\
}
% make the title area
\maketitle \thispagestyle{empty} \vspace{-15mm}
{\renewcommand{\baselinestretch}{1}
\begin{abstract}

In this paper we consider the secure transmission over the fast fading multiple antenna Gaussian broadcast channels with confidential messages (FMGBC-CM), where a multiple-antenna transmitter sends independent confidential messages to two users with information theoretic secrecy and only the statistics of the receivers' channel state information are known at the transmitter. We first use the same marginal property of the FMGBC-CM to classify the non-trivial cases, i.e., those not degraded to the common wiretap channels. We then derive the achievable rate region for the FMGBC-CM by solving the channel input covariance matrices and the inflation factor. Due to the complicated rate region formulae, we resort to low SNR analysis to investigate the characteristics of the channel. Finally, the numerical examples show that under the information-theoretic secrecy requirement both users can achieve positive rates simultaneously.
\end{abstract}

\section{ Introduction}
Traditionally, the security of data transmission has been ensured
by the key-based enciphering. However, for secure communication in
large-scale wireless networks, the key distributions and
managements may be challenging tasks
\cite{Liang_same_marginal}\cite{Secureconnect}. The physical-layer
security introduced in
\cite{Wyner_wiretap}\cite{csiszar1978broadcast} is appealing due
to its keyless nature. One of the fundamental setting for the
physical-layer security is the wiretap channel. In this channel, the transmitter wishes to send messages securely to a legitimate
receiver and to keep the eavesdropper as ignorant of the message
as possible. Wyner first characterized the secrecy capacity of the
discrete memoryless wiretap channel \cite{Wyner_wiretap}. The
secrecy capacity is the largest rate communicated between the
source and legitimate receiver with the eavesdropper knowing no
information of the messages. Motivated by the demand of high data
rate transmission, the multiple antenna systems with security
concern were considered by several works. In \cite{Shafiee_secrecy_2_2_1_J}, Shafiee and Ulukus
first proved the secrecy capacity of a Gaussian channel with two-input, two-output,
single-antenna-eavesdropper. Then the authors of
\cite{Khisti_MIMOME,Oggier_MIMOME, Liu_MIMO_wiretap} extended the
secrecy capacity result to the Gaussian multiple-input
multiple-output, multiple-antenna-eavesdropper channel. On
the other hand, the impacts of fading channels on the secure transmission were
considered in \cite{Liang_fading_secrecy}. Note that \cite{Shafiee_secrecy_2_2_1_J,Khisti_MIMOME,Oggier_MIMOME,
Liu_MIMO_wiretap,Liang_fading_secrecy} require full channel state
information at the transmitter (CSIT). When there is only partial CSIT,
several works considered the secure transmission under this condition \cite{Khisti_MISOME,Goel_AN,Pulu_ergodic,Li_fading_secrecy_j,
gopala2008secrecy}. The artificial noise (AN) assisted secure
beamforming is a promising technique for the partial CSIT cases, where
in addition to the message-bearing signal, an AN is intentionally
transmitted to disrupt the eavesdropper's reception
\cite{Khisti_MISOME}\cite{Goel_AN}.  Indeed, adding AN in transmission is
crucial in increasing the secrecy rate in fading wiretap channels. However, the covariance
matrices of AN in \cite{Goel_AN}\cite{Khisti_MISOME} is
heuristically selected without optimization, and the resulting
secrecy rate is not optimal. In \cite{Pulu_ergodic}, the secure transmission under fast fading channels with only statistical CSIT and without AN is considered.
Although the secrecy capacity for single antenna system with
partial CSIT was found in \cite{gopala2008secrecy}, the decoding
latency of the transmission scheme proposed in
\cite{gopala2008secrecy} is much longer than the common fast fading channels, e.g., \cite{Khisti_MISOME,Goel_AN,Pulu_ergodic,Li_fading_secrecy_j}, and may be unacceptable in practice.

However, the assumptions of wiretap channels with full or partial CSIT may not be practical. That is, the eavesdroppers needs to feedback the perfect/statistical CSI to transmitter or the transmitter needs to know this CSI by some means. On the contrary, the eavesdroppers may not be motivated to feedback this information. Furthermore, the eavesdroppers may feedback the wrong CSI to destroy the secure transmission. Thus in this paper, we consider the multiple antenna Gaussian broadcast channel with confidential messages (MGBC-CM) \cite{Liu_MISO} under fast fading channels (abbreviated as \textit{FMGBC-CM}). In the FMGBC-CM, both receivers are legitimate users such that they both are willing to feedback accurate CSI to maintain their secure transmission, and not to be eavesdropped by the other user. In the considered FMGBC-CM, we
assume that the transmitter only has the statistics of the channels from both receivers. This is to taking the practical issues into account, such as the limited bandwidth of the feedback channels or the speed of the channel estimation at the receivers. And to the best knowledge of the authors, this problem has not been considered in the literature.

The main contribution of this paper is to provide an achievable rate region with explicit channel input covariance matrices of both users. An iterative algorithm is proposed to solve the inflation factor of the linear assignment Gelfand-Pinsker coding (LA-GPC) \cite{Gelfand_Pinsker} used in the adopted transmission scheme. To accomplish these, we first classify the non-trivial cases such that the FMGBC-CM is not degraded as the conventional wiretap channel, i.e., both users have positive secure transmission rates, by the same marginal property. We then prove that the MISO GBC-CM with identical and independently distributed (i.i.d.) Rayleigh fading channels is degraded as the MISO Gaussian wiretap channel. Thus in this paper we consider the non-i.i.d. Rayleigh fading MISO and Rician fading MISO BC-CM, respectively.

The rest of the paper is organized as follows. In Section
\ref{Sec_system_model} we introduce the considered system model. We then provide the necessary conditions for the FMGBC-CM to be not degraded as a conventional wiretap channel in Section \ref{sec_necessary_condition}.
In Section \ref{Sec_rate_region_MGBC_CM}, we derive the achievable secrecy rate region of the FMGBC-CM. An achievable selection of the channel input covariance matrices and an iterative scheme for solving the inflation factor are also provided to calculate the explicit rate region. In Section \ref{Sec_low_SNR}, we demonstrate the secrecy rate region in low SNR regime and the optimal signaling. In Section \ref{Sec_simulation} we illustrate the
numerical results. Finally, Section \ref{Sec_conclusion}
concludes this paper.

\section{System model}\label{Sec_system_model}

In this paper we consider the FMGBC-CM system as shown in Fig. \ref{Fig_sys},
where the transmitter has $n_T$ antennas and the receiver 1 and 2 each has single antenna. The received
signals at the two receivers can be respectively represented as \footnote{In this paper, lower and upper case bold alphabets denote vectors
and matrices, respectively. Italic upper alphabets with and without boldface denote random vectors and variables, respectively. The
$i$th element of vector $\ba$ is denoted by $a_i$. The
superscript $(.)^H$ denotes the transpose complex conjugate.
$|\mathbf{A}|$ and $|a|$ represent the determinant of the square
matrix $\mathbf{A}$ and the absolute value of the scalar variable
$a$, respectively. A diagonal matrix whose diagonal entries are
$a_1 \ldots a_k$ is denoted by $diag(a_1 \ldots a_k)$. The trace
of $\mathbf{A}$ is denoted by $\tr(\mathbf{A})$. We define $C(x)
\triangleq \log(1+x)$ and $(x)^+\triangleq\max\{0,\,x\}$. The
mutual information between two random variables is denoted by
$I(;)$. $\mathbf{I}_n$ denotes the $n$ by $n$ identity matrix.
$\mathbf{A}\succ 0$ and $\mathbf{A}\succeq 0$ denote that
$\mathbf{A}$ is a positive definite and positive semi-definite
matrix, respectively.}
\begin{align}
Y_{1,k}&={\bm{H}_{1,k}}^H\bm{X}_k+N_{1,k}, \label{EQ_main}\\
Y_{2,k}&={\bm{H}_{2,k}}^H\bm{X}_k+N_{2,k}\label{EQ_Eve},
\end{align}
where $\bm{X}_k\in\mathds{C}^{n_T\times 1}$ is the transmit
vector, $k$ is the time index, respectively, which denote the fading vector channels from
transmitter to the two receivers, and
$N_{1,k}$ and $N_{2,k}$ are circularly symmetric complex additive
white Gaussian noises with variances one at receiver 1 and receiver 2,
respectively. In this system, we assume that only the statistics of both channels are known at transmitter, to take the practical issues of system design into account, such as the limited bandwidth of the feedback channels or the speed of the channel estimation at the receivers. We also assume that the receiver 1 and 2 perfectly know their channel vectors $\bm{H}_1$ and $\bm{H}_2$, respectively.
Without loss of
generality, in the following we omit the time index to simplify
the notation. We consider the power constraint as
\[
\mbox{tr}(E[\bm{X}\bm{X}^H])\leq P_T.
\]

The perfect secrecy and secrecy capacity are defined as follows.
Consider a $(2^{nR_1}, 2^{nR_2},n)$-code with an encoder that maps the
message $W_1\in \calW_1=\{1,2,\ldots, 2^{nR_1}\}$ and $W_2\in\calW_2=\{1,2,\ldots, 2^{nR_2}\}$ into a length-$n$
codeword, and receiver 1 and receiver 2 map the
received sequence $Y_1^n$ and $Y_2^n$ (the collections of $Y_1$ and $Y_2$, respectively, over code length $n$) from the MISO channel to the estimated
message $\hat W_1\in\calW_1$ and $\hat W_2\in\calW_2$, respectively. Since $\bm{H}_1$ and $\bm{H}_2$ are both known at receiver 1 and 2, respectively, we can treat them as the channel outputs similar to \cite{caire_channel_with_SI}. We then have the following definition of secrecy capacity.

\begin{definition}[Secrecy capacity region] \label{Def_Perfect}
 {\it Perfect secrecy with rate pair $(R_1,\,R_2)$
is achievable if, for any positive $\varepsilon$ and $\varepsilon'$, there
exists a sequence of $(2^{nR_1},2^{nR_2}, n)$-codes and an integer $n_0$ such
that for any $n>n_0$
\begin{align}\label{eq_equivocation_given_h}
&I(W_1;Y_2^n,\bm{H}_2^n)/n<\varepsilon, \mathrm{and
}\;\;{\rm Pr}(\hat{W}_1\neq W_1) \leq \varepsilon',\\
&I(W_2;Y_1^n,\bm{H}_1^n)/n<\varepsilon, \mathrm{and
}\;\;{\rm Pr}(\hat{W}_2\neq W_2) \leq \varepsilon',
\end{align}
where $\bm{H}_1^n$ and $\bm{H}_2^n$ are the collections of $\bm{H}_1$ and
$\bm{H}_2$ over code length $n$, respectively. The {\rm secrecy capacity region} is the closure of the set of all achievable
rate pairs $(R_1,\,R_2)$.}
\end{definition}

Note that as shown in the footnote, italic upper alphabets with and without boldface denote random vectors and variables, respectively. By treating $\bm{H}_1$ and $\bm{H}_2$ as the channel outputs, we can extend the achievable rate region of the discrete memoryless MBC-CM from \cite{Liu_MISO} as
\[
(R_1,R_2)\in \mbox{co}\left\{\underset{\varpi\in\Omega}\bigcup \mathcal{R}_I(\varpi)\right\},
\]
where co$\{.\}$ denotes the convex closure; $\mathcal{R}_{I}(\varpi)$ denotes the union of all $(R_1,R_2)$ satisfying
\begin{align}
R_1 &\leq (I(\bm{V}_1;Y_1,\bm{H}_1)-I(\bm{V}_1;Y_2,\bm{H}_2,\bm{V}_2))^+,\label{EQ_R1_DMC}\\
R_2 &\leq (I(\bm{V}_2;Y_2,\bm{H}_2)-I(\bm{V}_2;Y_1,\bm{H}_1,\bm{V}_1))^+,\label{EQ_R2_DMC}
\end{align}
for any given joint probability density $\varpi$ belonging to the class of joint probability densities $p(\bm{v}_1,\bm{v}_2,\bm{x},y_1,y_2,\bm{h}_1,\bm{h}_2)$, denoted by  $\Omega$, that factor as $p(\bm{v}_1,\bm{v}_2)p(\bm{x}|\bm{v}_1,\bm{v}_2)p(y_1,y_2,\bm{h}_1,\bm{h}_2|\bm{x})$; $\bm{V}_1$ and $\bm{V}_2$ are the auxiliary random vectors for user 1 and 2, respectively.\\

Note that we can further rearrange the right hand side (RHS) of \eqref{EQ_R1_DMC} as
\begin{align}
R_1 \overset{(a)}\leq &(I(\bm{V}_1;Y_1,\bm{H}_1)-I(\bm{V}_1;\bm{V}_2,\bm{H}_2)-I(\bm{V}_1;Y_2|\bm{V}_2,\bm{H}_2))^+\notag\\
\overset{(b)}=&(I(\bm{V}_1;Y_1,\bm{H}_1)-I(\bm{V}_1;\bm{V}_2)-I(\bm{V}_1;Y_2|\bm{V}_2,\bm{H}_2))^+\notag\\
\overset{(c)}=&(I(\bm{V}_1;Y_1|\bm{H}_1)+I(\bm{V}_1;\bm{H}_1)-I(\bm{V}_1;Y_2|\bm{V}_2,\bm{H}_2)-I(\bm{V}_1;\bm{V}_2))^+\notag\\
\overset{(d)}=&(I(\bm{V}_1;Y_1|\bm{H}_1)-I(\bm{V}_1;Y_2|\bm{V}_2,\bm{H}_2)-I(\bm{V}_1;\bm{V}_2))^+,\label{EQ_R1_DMC2}
\end{align}
where (a) is by applying the chain rule of mutual information to the second term on the RHS of \eqref{EQ_R1_DMC}; (b) is due to $\bm{V}_1$ and $\bm{V}_2$ are independent of $\bm{H}_2$; (c) is again applying the chain rule to the first term; (d) is due to the fact that there is only statistical CSIT, and $\bm{V}_1$ is independent of $\bm{H}_1$. Thus $I(\bm{V}_1;\bm{H}_1)=0$. Similarly, we can can rearrange $R_{2}$ as
\begin{equation}
R_2 \leq (I(\bm{V}_2;Y_2|\bm{H}_2)-I(\bm{V}_2;Y_1|\bm{V}_1,\bm{H}_1)-I(\bm{V}_1;\bm{V}_2))^+.\label{EQ_R2_DMC2}
\end{equation}
\section{Conditions for non-degraded FMGBC-CM}\label{sec_necessary_condition}
Before investigating the rate region of the FMGBC-CM, we need to exclude the cases that FMGBC-CM are degraded as fast fading wiretap channels, for which one of the two receivers always has zero rate. The capacity result can be concluded according to the results in \cite{SC_TIFS}. Furthermore, from \cite{SC_TIFS} we know that for such channel the optimal eigenvectors and eigenvalues of the channel input covariance matrix are arbitrarily orthonormal basis and uniform allocated powers, respectively. \\

Our first result is as following.\\
\begin{lemma}\label{Lemma_necessary_conditon_FMGBCCM}
A necessary condition for the two users in the fast Rayleigh FMGBC-CM both having positive rates is that $\bm{H}_1$ and $\bm{H}_2$ are not i.i.d.. \\
\end{lemma}

Note that the wiretap channel is a special case of the GBC-CM which can be easily derived by letting $W_2$ in the GBC-CM as null. To prove this result, we need to introduce the following lemma first, which is extended from \cite[Lemma 4]{Liu_MISO}\\
\begin{lemma}\label{Def_same_marginal}
Let $\mathcal{P}$ denote the set of channels $p(\tilde{y}_1,\,\tilde{y}_2,\,\tilde{\bm{h}}_1,\,\tilde{\bm{h}}_2|\bm{x})$ whose marginal distributions satisfy
\begin{align}
&p_{\tilde{Y}_1,\tilde{\bm{H}}_1|\bm{X}}(\tilde{y}_1,\tilde{\bm{h}}_1|\bm{x})=p_{Y_1,\bm{H}_1|\bm{X}}(y_1,\bm{h}_1|\bm{x}),\\
&p_{\tilde{Y}_2,\tilde{\bm{H}}_2|\bm{X}}(\tilde{y}_2,\tilde{\bm{h}}_2|\bm{x})=p_{Y_2,\bm{H}_2|\bm{X}}(y_2,\bm{h}_2|\bm{x}),
\end{align}
for all $y_1,\,y_2,\, \mbox{and } \bm{x}$. The secrecy capacity region is the same for all channels $p(\tilde{y}_1,\,\tilde{y}_2,\,\tilde{\bm{h}}_1,\,\tilde{\bm{h}}_2|\bm{x})\in\mathcal{P}$.\\
\end{lemma}

Note that $p(\tilde{y}_1,\,\tilde{y}_2,\,\tilde{\bm{h}}_1,\,\tilde{\bm{h}}_2|\bm{x})$ is from the factorization below \eqref{EQ_R2_DMC}. Due to limited space, we give a sketch of the proof of Lemma \ref{Lemma_necessary_conditon_FMGBCCM} in the following. Assume both channels are i.i.d., i.e., $\bm{H}_1\sim CN(0,\sigma_1^2\bI)$ and $\bm{H}_2\sim CN(0,\sigma_2^2\bI)$. With the same marginal property in Definition \ref{Def_same_marginal}, we can replace $\bm{H}_1$ in \eqref{EQ_main} by $(\sigma_1/\sigma_2)\bm{H}_2$ without affecting the capacity. Thus we have a  new pair of channels with the same capacity as \eqref{EQ_main} and \eqref{EQ_Eve}
\begin{align}
Y_1'&=(\sigma_1/\sigma_2)\bm{H}_2^H\bm{X}+N_1,\notag\\
Y_2&=\bm{H}_2^H\bm{X}+N_2,\notag
\end{align}
which can be further represented as
\begin{align}
Y_1''&=\bm{H}_2^H\bm{X}+(\sigma_2/\sigma_1)N_1,\notag\\
Y_2&=\bm{H}_2^H\bm{X}+N_2.\notag
\end{align}
Thus as long as $\sigma_1>\sigma_2$, we can have the Markov chain $\bm{X}\rightarrow Y_1''\rightarrow Y_2$. On the other hand, by extending the outer bound of \cite[Theorem 3]{Liu_MISO}, we know that \textit{less noisy} \cite[Ch. 5]{Kim_lecture} makes one of the FMGBC-CM user have zero rate. Since $Y_2$ is degraded of $Y_1''$, and degradedness is more strict than the less noisy, thus we know $R_2=0$. Similarly, when $\sigma_2>\sigma_1$, we know that $R_1=0$.\\

An intuitive explanation is that, if a message can be successfully decoded by the inferior user, then the superior user is also ensured of decoding it. Thus the secrecy rate of the degraded user is zero. Based on the concept mentioned above, we can extend Lemma \ref{Lemma_necessary_conditon_FMGBCCM} to the following.\\

\begin{coro}
A necessary condition for the two users in the fast Rayleigh FMGBC-CM both having positive rates is that the covariance matrices of $\bm{H}_1$ and $\bm{H}_2$ should not be scaled of each other. \\
\end{coro}

Therefore to avoid the investigation of such cases, in the following we assume $\bm{H}_1\sim CN(\bm\mu_1,\bK_{\bm{H}_1})$ and $\bm{H}_2\sim CN(\bm\mu_2,\bK_{\bm{H}_2})$, where $\bK_{\bm{H}_1}$ and $\bK_{\bm{H}_2}$ may not be scaled of each other.\\

Two special cases with single input single output (SISO) antenna GBC-CM are also summarized as follows.\\

\begin{coro}
All SISO Rayleigh fading GBC-CMs with only statistical CSIT degrade as wiretap channels.\\
\end{coro}

\begin{coro}
All SISO Rician fading GBC-CMs with only statistical CSIT degrade as wiretap channels if the channels have the same $K$-factor.\\
\end{coro}

\textit{Remark:} Directly verifying the \textit{less noisy} \cite{Kim_lecture} property, i.e., $I(\bm{V},Y_1)\lessgtr I(\bm{V},Y_2)$ from extending the upper bound derivation in \cite[Theorem 3, Example 1]{Liu_SISO} to multiple antenna case involves manipulations of two rates similar to \eqref{EQ_R1_GBC} and \eqref{EQ_R2_GBC}, which is intractable. $\bm{V}$ here can be either $\bm{V}_1$ or $\bm{V}_2$.

\section{The achievable secrecy rate region of FMGBC-CM}\label{Sec_rate_region_MGBC_CM}

Due to the fact that there is only statistical CSIT, we can not use the original minimum mean square error (MMSE) inflation factor as Costa \cite{CostaDPC}, where the exact channel state information is required. Thus we need to re-derive the achievable rate region of the FMGBC-CM instead of directly using Liu's result in \cite[Lemma 3]{Liu_MISO}. To derive the new achievable rate region, we resort to the linear assignment Gel'fand Pinsker coding \cite{Gelfand_Pinsker}, which is the generalized case of DPC, to deal with the fading channels, similar to our previous work \cite{pslin_CR}. For the FMGBC-CM, we consider the secret LA-GPC with Gaussian codebooks. First, separate the channel input $\bm{X}$ into two random vectors $\bm{U}_1$ and $\bm{U}_2$ so that $\bm{X}=\bm{U}_1+\bm{U}_2$. Then $\bm{U}_1$ and $\bm{U}_2$ are chosen as follows:
\begin{align}
&\bm{U}_1\sim CN(\mathbf{0},\mathbf{K}_{\bm{U}_1}),\\
&\bm{U}_2\sim CN(\mathbf{0},\mathbf{K}_{\bm{U}_2}),
\end{align}
where $\bm{U}_2$ is independent of $\bm{U}_1$, $\mathbf{K}_{\bm{U}_1}\succeq 0$ and $\mathbf{K}_{\bm{U}_2}\succeq 0$ are the covariance matrices of $\bm{U}_1$ and $\bm{U}_2$, respectively. After that, we do the decomposition $\mathbf{K}_{\bm{U}_1}=\mathbf{T}_1 \mathbf{T}_1^H$, and define $\bm{U}_1' \sim CN(\mathbf{0},\mathbf{I}_N)$ so that $\bm{U}_1=\mathbf{T}_1\bm{U}_1'$, where $\bT_1\in\mathds{C}^{n_T\times N}$ and $N$ is the rank of $\mathbf{K}_{\bm{U}_1}$.
The auxiliary random variables are then defined as:
\begin{align}
&\bm{V}_1=\bm{U}_1'+\ba\bm{H}_1^H\bm{U}_2, \label{dpc}\\
&\bm{V}_2=\bm{U}_2,
\end{align}
where $\mathbf{a}$ is the inflation factor in LA-GPC. The reason to choose \eqref{dpc} is that if we do LA-GPC for $\bm{U}_1$ directly, i.e., $\bm{V}_1=\bm{U}_1+\ba\bm{H}_1^H\bm{U}_2$, after substituting it into the RHS of \eqref{EQ_R1_DMC} and \eqref{EQ_R2_DMC2}, we can find that the rate formula includes $\log|\mathbf{K}_{\bm{U}_1}|$ when calculating $I(\bm{V}_1;\bm{V}_2)$, which requires $\mathbf{K}_{\bm{U}_1}\succ 0$. However, the expression of \eqref{dpc} would bypass this constraint.
Note that in the rest of this paper, for convenience of computation, we combine $\ba\bm{H}_1^H$ as $\mathbf{b}$. To present the rate regions compactly, recall that the permutation $\pi$ specifies the encoding order, i.e., the message of user $\pi_1$ is encoded first while the message of user $\pi_2$ is encoded second.\\

\begin{lemma} \label{Lemma_rate_region}
 Let $\mathcal{R}(\mathbf{K}_{\bm{U}_{\pi_1}},\mathbf{K}_{\bm{U}_{\pi_2}})$ denote the union of all $(R_{\pi_1},R_{\pi_2})$ satisfying
\begin{align}
R_{\pi_1} &\leq \left( E_{\bm{H}_{\pi_1}}[\log(1+{\bm{H}_{\pi_1}}^H(\mathbf{K}_{\bm{U}_{\pi_1}}+\mathbf{K}_{\bm{U}_{\pi_2}}){\bm{H}_{\pi_1}})]-\bigtriangleup\right)^+,\label{EQ_R1_GBC}\\
R_{\pi_2} &\leq \left(  E_{\bm{H}_{\pi_2}}[\log(1+{\bm{H}_{\pi_2}}^H(\mathbf{K}_{\bm{U}_{\pi_1}}+\mathbf{K}_{\bm{U}_{\pi_2}}){\bm{H}_{\pi_2}})]-\bigtriangleup\right)^+\label{EQ_R2_GBC},
\end{align}
where
\begin{align}
&\bigtriangleup\triangleq E_{\bm{H}_{\pi_2}}[\log(1+{\bm{H}_{\pi_2}}^H\mathbf{K}_{\bm{U}_{\pi_1}}{\bm{H}_{\pi_2}})]+\notag\\
&E_{\bm{H}_{\pi_1}}\!\!\!\!\left[\log\!\! \left|\!\!\!\! \begin{array}{cc}
                                                                                                   { \mathbf{I}+\mathbf{bK}_{\mathbf{U}_{\pi_2}}\mathbf{b}^H }&{ (\mathbf{T}_1^H+\mathbf{bK}_{\mathbf{U}_{\pi_2}}){\bm{H}_{\pi_1}} }  \\
                                                                                                   { {\bm{H}_{\pi_1}}^H(\mathbf{T}_1+\mathbf{K}_{\bm{U}_{\pi_2}}\mathbf{b}^H) }&{ 1+{\bm{H}_{\pi_1}}^H(\mathbf{K}_{\bm{U}_{\pi_1}}+\mathbf{K}_{\bm{U}_{\pi_2}}){\bm{H}_{\pi_1}} }
                                                                                                  \end{array}\label{EQ_M}
\!\!\!\!\!\!    \right|   \right].
\end{align}
Then any rate pair
\[
(R_1,R_2)\in \mbox{co}\left\{\underset{\mbox{tr}(\mathbf{K}_{\bm{U}_{\pi_1}}+\mathbf{K}_{\bm{U}_{\pi_2}})\leq P_T}\bigcup\mathcal{R}(\mathbf{K}_{\bm{U}_{\pi_1}},\mathbf{K}_{\bm{U}_{\pi_2}})\right\}
\]
is achievable for the FMGBC-CM.\\
\end{lemma}

Due to the limited space, we do not provide the derivation in detail. In the following, we provide an achievable scheme to approximately achieve the above two bounds in \eqref{EQ_R1_GBC} and \eqref{EQ_R2_GBC}.\\

\begin{lemma}\label{lemma_optimal_input_cov_mat}
With the selection $\mathbf{K}_{\bm{U}_{\pi_1}}^*=\alpha P_T\mathbf{e}_{\pi_1}^*(\mathbf{e}_{\pi_1}^*)^H$ and $\mathbf{K}_{\bm{U}_{\pi_2}}^*=(1-\alpha) P_T\mathbf{e}_{\pi_2}^*(\mathbf{e}_{\pi_2}^*)^H$, where $||\mathbf{e}_{\pi_1}^*||^2=1$, $||\mathbf{e}_{\pi_2}^*||^2=1$, and
\begin{align}
\mathbf{e}_{\pi_1}^*&=\max_{\mathbf{e}_{\pi_1}}\frac{\mathbf{e}_{\pi_1}^H\left(\mathbf{I}+\alpha P_T\left(\bK_{{\bm{H}}_{\pi_1}}+\bm\mu_{\pi_1}\bm\mu_{\pi_1}^H\right)\right)\mathbf{e}_{\pi_1}}{\mathbf{e}_{\pi_1}^H\left(\mathbf{I}+\alpha P_T\left(\bK_{{\bm{H}}_{\pi_1}}+\bm\mu_{\pi_2}\bm\mu_{\pi_2}^H\right)\right)\mathbf{e}_{\pi_1}},\label{EQ_e1}
\\
\mathbf{e}_{\pi_2}^*&=\max_{\mathbf{e}_{\pi_2}}\frac{\mathbf{e}_{\pi_2}^H\left(\mathbf{I}+\frac{(1-\alpha)P_T\left(\bK_{{\bm{H}}_{\pi_2}}+\bm\mu_{\pi_2}\bm\mu_{\pi_2}^H\right)}{1+\alpha P_T(\mathbf{e}_{\pi_1}^*)^H\left(\bK_{{\bm{H}}_{\pi_2}}+\bm\mu_{\pi_2}\bm\mu_{\pi_2}^H\right)\mathbf{e}_{\pi_1}^*}\right)
\mathbf{e}_{\pi_2}}{\mathbf{e}_{\pi_2}^H\left(\mathbf{I}+\frac{(1-\alpha)P_T\left(\bK_{{\bm{H}}_{\pi_1}}+\bm\mu_{\pi_1}\bm\mu_{\pi_1}^H\right)}{1+\alpha P_T(\mathbf{e}_{\pi_1}^*)^H\left(\bK_{{\bm{H}}_{\pi_1}}+\bm\mu_{\pi_1}\bm\mu_{\pi_1}^H\right)\mathbf{e}_{\pi_1}^*}\right)\mathbf{e}_{\pi_2}},\label{EQ_e2}
\end{align}
where $\alpha$ is the ratio of power allocated to user $\pi_1$, we can get the non-trivial rate region for the FMGBC-CM as
\[
(R_1,R_2)\in \mbox{co}\left\{\underset{0\leq\alpha\leq 1}\bigcup\mathcal{R}(\mathbf{K}_{\bm{U}_{\pi_1}}^*,\mathbf{K}_{\bm{U}_{\pi_2}}^*)\right\}.
\]
\end{lemma}
\vspace{0.5cm}
Due to the limited space, only the proof sketch is given as follows. Instead of solving $\bK_{\bm{U}_{\pi_1}}$ and $\bK_{\bm{U}_{\pi_2}}$ from \eqref{EQ_R1_GBC} and \eqref{EQ_R2_GBC} directly, which may be intractable, we resort to solving the upper bound of the rate region described by Lemma \ref{Lemma_rate_region}. That is, the transmitter can use full CSIT to design the inflation factor. Then it is clear that the optimal $\bb$ is the MMSE estimator
\begin{equation}\label{EQ_MMSE_b}
\bb=\bT_{\pi_1}^H\bm{H}_{\pi_1}\bm{H}_{\pi_1}^H/(1+\bm{H}_{\pi_1}^H\bK_{\bm{U}_{\pi_1}}\bm{H}_{\pi_1}).
 \end{equation}
Then after some manipulations and applying the Jensen's inequality followed by the unit rank selection of $\bK_{\bm{U}_{\pi_1}}$ and $\bK_{\bm{U}_{\pi_2}}$, we can have the Rayleigh quotient form as \eqref{EQ_e1} and \eqref{EQ_e2}. Note that with \cite[Property 2, 3]{Petropulu_MIMOME} it can be proved that when the number of transmit antenna is 2 with $\bK_{\bm{H}_{\pi_1}}-\bK_{\bm{H}_{\pi_2}}\nsucceq 0$, then unit rank $\bK_{\bm{U}_{\pi_1}}$ and $\bK_{\bm{U}_{\pi_2}}$ is optimal for the considered upper bound.\\

 After deriving the covariance matrices, we then need to solve the inflation factor due to the fact that there is indeed no full CSIT. Here we resort to the following fixed point iteration to solve $\bb$
\begin{align}\label{EQ_iterative_b}
& \mathbf{b}=-(E_{\bm{H}_1}[\mathbf{A}_1^H])^{-1}E_{\bm{H}_1}[\mathbf{A}_2^H\bm{H}_1^H]\triangleq f(\bb),
 \left[
                                                                                                            \begin{array}{c}
                                                                                                              \mathbf{A}_1 \\
                                                                                                              \mathbf{A}_2 \\
                                                                                                            \end{array}
                                                                                                          \right]\triangleq\mathbf{M}^{-1}\left[
                                                                                                                    \begin{array}{c}
                                                                                                                     \mathbf{I} \\
                                                                                                                      \mathbf{0} \\
                                                                                                                    \end{array}
                                                                                                                  \right],
\end{align}
where $\bM$ is defined as the block matrix inside the determinant of the second term in \eqref{EQ_M}.
Note that \eqref{EQ_iterative_b} is derived by $\partial R_1/\partial \bb=0$. Note also that the iteration stops when the maximum relative error of $R_1$ and $R_2$ in the successive iterations is less than a predefined value. The iteration steps are summarized in Table \ref{TA iterative steps}.

\begin{table} [ht]
\begin{center}
\caption{The iterative steps for solving $\bb$.}
\begin{tabular}{l l} \label{TA iterative steps}
Step 1 & Set $i=0$ and initialize $\mathbf{b}^{(i)}=\mathbf{0}$. Also initialize $\be_{\pi_1}$ and $\be_{\pi_2}$ as \notag\\
&\eqref{EQ_e1} and \eqref{EQ_e2}, respectively.\\
Step 2 & Evaluate $\mathbf{b}^{(i+1)}=f(\mathbf{b}^{(i)})$.\\
Step 3 & Let $i=i+1$ and repeat Step 2 until\notag\\
& $\max\{R_1^{(i)}-R_1^{(i-1)},\,R_2^{(i)}-R_2^{(i-1)}\}<\varepsilon$.
\end{tabular}
\end{center}
\end{table}

\section{ Low SNR Analysis}\label{Sec_low_SNR}
In this section we study the achievable secrecy rate region in the low-SNR regime. Note that operation at low SNRs is beneficial from a security perspective since it is generally difficult for a eavesdropper to detect the signal \cite{Gursoy_security_low_SNR}. In addition, due to the rate region in Lemma \ref{Lemma_rate_region} is complicated to analyze, we resort to the low SNR regime to get some insights.\\

\begin{lemma}
In the low SNR regime, the optimal input covariance matrices $\mathbf{K}_{\bm{U}_1}$ and $\mathbf{K}_{\bm{U}_2}$ are both unit rank, with the direction aligned to the eigenvector corresponding to the maximum eigenvalue of $\mathbf{K}_{\bm{H}_1}-\mathbf{K}_{\bm{H}_2}$ and $\mathbf{K}_{\bm{H}_2}-\mathbf{K}_{\bm{H}_1}$, respectively. And the asymptote of the secrecy rate region is
\begin{align}
R_1 &\leq \left(\frac{\alpha P_T}{\ln 2}\lambda_{max}(\mathbf{K}_{\bm{H}_1}-\mathbf{K}_{\bm{H}_2})\right)^+, \label{R1_low_SNR}\\
R_2 &\leq \left(\frac{(1-\alpha) P_T}{\ln 2}\lambda_{max}(\mathbf{K}_{\bm{H}_2}-\mathbf{K}_{\bm{H}_1})\right)^+.\label{R2_low_SNR}\\\notag
\end{align}
\end{lemma}
Note that the unit rank result is consistent to that of MGBC-CM with perfect CSIT, also our selection of the $\bK_{\bm{U}_{\pi_1}}^*$ and $\bK_{\bm{U}_{\pi_2}}^*$ in Lemma \ref{lemma_optimal_input_cov_mat}.
From the rate region described in \eqref{R1_low_SNR} and \eqref{R2_low_SNR}, we have\\
\begin{coro}\label{coro_low_snr_positive_rates_condition}
In the low SNR regime, both users can have positive rates simultaneously if and only if $\mathbf{K}_{\bm{H}_1}-\mathbf{K}_{\bm{H}_2}$ is indefinite.\\
\end{coro}

\textit{Remark:} Note that it can be easily seen that Corollary \ref{coro_low_snr_positive_rates_condition} includes Lemma \ref{Lemma_necessary_conditon_FMGBCCM} in the low SNR regime. That is, if both $\bm{H}_1$ and $\bm{H}_2$ are i.i.d., respectively, then $\mathbf{K}_{\bm{H}_1}-\mathbf{K}_{\bm{H}_2}$ has all eigenvalues positive or negative. Thus the two results coincide.

\section{Numerical Results}\label{Sec_simulation}
In this section, we compare the rate regions of our proposed achievable scheme under both Rayleigh (with at least one channel having non-i.i.d. distribution) and Rician fading channels to that of  full CSIT MGBC-CM, respectively. We set $n_T=2$ and the power constraint $P_T=10$, respectively. We also set the stopping criteria of the iterative algorithm as $\varepsilon=10^{-3}$. In the simulation of Rayleigh fading case, we set the covariance matrices of the two channels as
\begin{equation}\label{EQ_Kh_example}
\bK_{\bm{H}_1}=\left[\begin{array}{cc}
            0.2   &  0\\
             0  &  0.04
            \end{array}\right],\,\,\bK_{\bm{H}_2}=\left[\begin{array}{cc}
            0.1   &  0.08\\
             0.08  &  0.1
            \end{array}\right],
\end{equation}
which satisfy Lemma \ref{Lemma_necessary_conditon_FMGBCCM}. Since the selection of $\mathbf{K}_{\bm{U}_{\pi_1}}^*$ in Lemma \ref{lemma_optimal_input_cov_mat} is rank 1, we know that $\bb\in\mathds{C}^{2\times 1}$ by definition.
For the full CSIT case, we consider the rate region which is the convex closure of the following rate pair
\begin{align}
&R_{\pi_1}\leq E\left[\left(\log_{2}\frac{1+\bm{H}_{\pi_1}^H\bK_{\bm{U}_{\pi_1}}\bm{H}_{\pi_1}}{1+\bm{H}_{\pi_2}^H\bK_{\bm{U}_{\pi_1}}\bm{H}_{\pi_2}}\right)^+\right],\label{EQ_full_CSIT_R1}\\
&R_{\pi_2}\leq \notag\\ &E\left[\left(\log_{2}\frac{[1+\bm{H}_{\pi_2}^H(\bK_{\bm{U}_{\pi_1}}+\bK_{\bm{U}_{\pi_2}})\bm{H}_{\pi_2}](1+\bm{H}_{\pi_1}^H\bK_{\bm{U}_{\pi_1}}\bm{H}_{\pi_1})}{[1+\bm{H}_{\pi_1}^H(\bK_{\bm{U}_{\pi_1}}+\bK_{\bm{U}_{\pi_2}})\bm{H}_{\pi_1}](1+\bm{H}_{\pi_2}^H\bK_{\bm{U}_{\pi_1}}\bm{H}_{\pi_2})}\right)^+\right],\label{EQ_full_CSIT_R2}
\end{align}
with the power constraint $\mbox{tr}(\bK_{\bm{U}_{\pi_1}}+\bK_{\bm{U}_{\pi_2}})\leq 10$, where the optimal $\bK_{\bm{U}_{\pi_1}}$ and $\bK_{\bm{U}_{\pi_2}}$ are described in \cite[(16)]{Liu_MISO} and the optimal $\bb$ is as \eqref{EQ_MMSE_b}. Note that \eqref{EQ_full_CSIT_R1} and \eqref{EQ_full_CSIT_R2} are the straightforward extension of \cite{Liu_MISO} to the fast fading channels with full CSIT. From Fig. \ref{Fig_Rayleigh} we can easily see that the proposed transmission scheme for the fast FMGBC-CM with partial CSIT apparently outperforms the time sharing scheme. Time sharing means that the transmitter sends the two messages with different powers during a fraction of time where these powers satisfy the average power constraint. And in each fraction of time, the fast FMGBC-CM reduces to a fading Gaussian MISO wiretap channel. We also consider the case that the inflation factor $\bb$ uses the MMSE estimator with channel mean, i.e., $\bb=[0\,\,0]$, which is the same as treating interference as noise. From Fig. \ref{Fig_Rayleigh} it can be seen that the performance of treating interference as noise is slightly better than the time sharing. On the other hand, by comparing the regions of full and partial CSIT cases, we can easily find the impact of the CSIT to the rate performance.

For the Rician fading case, in addition to \eqref{EQ_Kh_example}, we let the mean vectors of $\bm{H}_1$ and $\bm{H}_2$ as
\[
\bm\mu_1=\left[\begin{array}{c}
            0.7  \\
             0.1
            \end{array}\right],\,\,\bm\mu_2=\left[\begin{array}{c}
            0.1  \\
             0.6
            \end{array}\right],
\]
respectively. From Fig. \ref{Fig_Rician},  we also can easily see that the CSIT plays an important role in improving the rate region. And time sharing is still the worst. With the aid of line of sight, the performances of all schemes under Rician fading are much better than the corresponding ones under Rayleigh fading. We also compare the case where $\bb$ is derived from substituting $\bm{H}_1=\bm\mu_1$ into \eqref{EQ_MMSE_b}. It can easily be seen that the proposed $\bb$ outperforms this selection of $\bb$. On the other hand, due to the gap is small, when low complexity is an important issue, the transmitter can choose such $\bb$ to implement the secure LA-GPC. Furthermore, we also show the rate region derived by $\bb=[0\,\,0]$. Similar to the Rayleigh fading case, this method is still worse than the proposed method, but slightly better than the time sharing. In Fig. \ref{Fig_SNR_Rayleigh} and Fig. \ref{Fig_SNR_Rician} we also compare the rate regions with different transmit SNRs under both Rayleigh and Rician fading channels. It can be seen that the rate regions of both cases enlarge with increasing transmit SNR.

\begin{figure}
\centering \epsfig{file=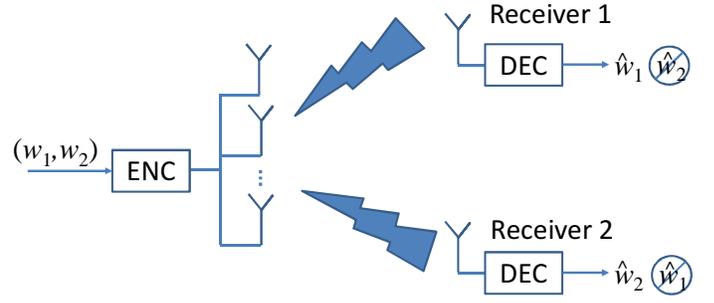 , width=0.5\textwidth}
\caption{The system model of FMGBC-CM.} \label{Fig_sys}
\end{figure}

\begin{figure}
\centering \epsfig{file=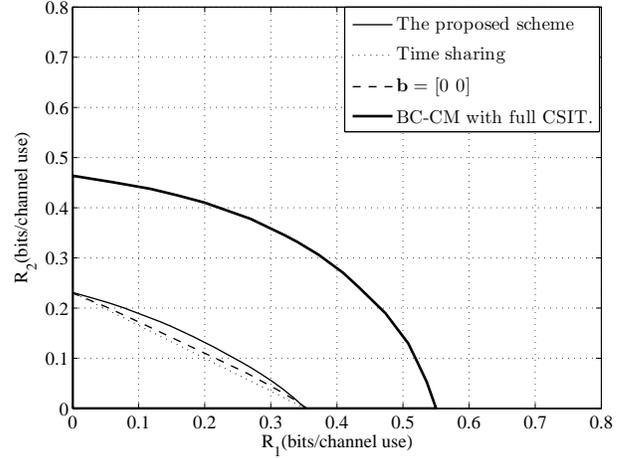 , width=0.5\textwidth}
\caption{The comparison of rate regions under fast Rayleigh fading channel with full and statistical CSIT.} \label{Fig_Rayleigh}
\end{figure}

\begin{figure}
\centering \epsfig{file=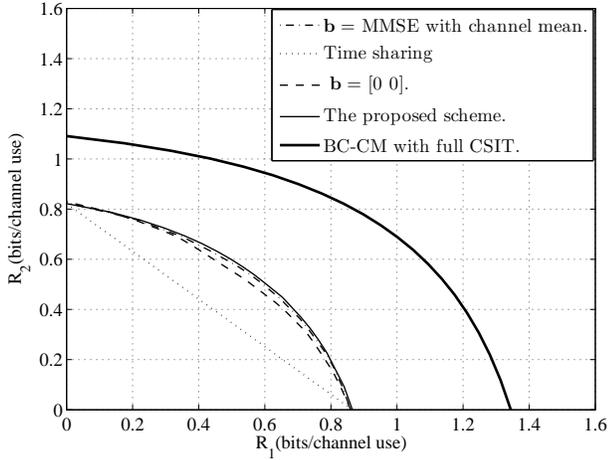 , width=0.5\textwidth}
\caption{The comparison of rate regions under fast Rician fading channel with full and statistical CSIT.} \label{Fig_Rician}
\end{figure}

\begin{figure}
\centering \epsfig{file=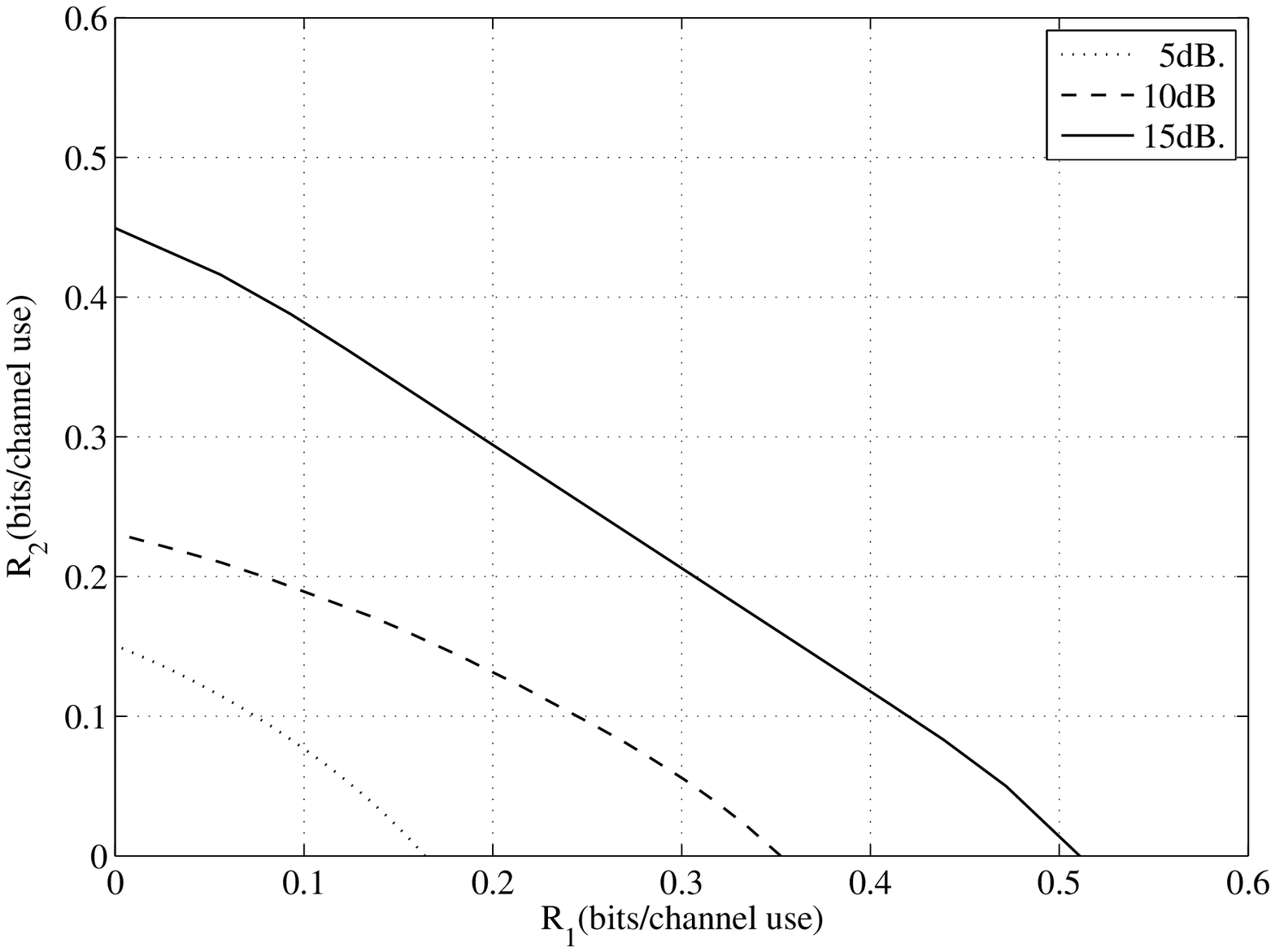 , width=0.5\textwidth}
\caption{The comparison of rate regions under fast Rayleigh fading channel with statistical CSIT and different transmit SNRs.} \label{Fig_SNR_Rayleigh}
\end{figure}

\begin{figure}
\centering \epsfig{file=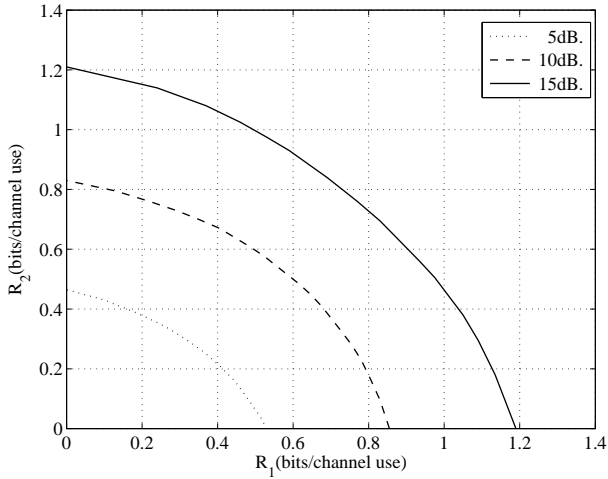 , width=0.5\textwidth}
\caption{The comparison of rate regions under fast Rician fading channel with statistical CSIT and different transmit SNRs.} \label{Fig_SNR_Rician}
\end{figure}
\section{Conclusion}\label{Sec_conclusion}
In this paper we considered the secure transmission over the fast fading multiple antenna Gaussian broadcast channels with confidential messages
 (FMGBC-CM), where a multiple-antenna transmitter sends independent confidential messages to two users with information theoretic secrecy
 and only the statistics of the receiver's channel state information are known at the transmitter. We first used the same marginal property of the
 FMGBC-CM to classify the non-trivial cases, i.e., not degraded to the common wiretap channels. We then derive the achievable rate region for
 the FMGBC-CM by solving the channel input covariance matrices and the inflation factor. We also provided a low SNR analysis for finding the
 asymptotic property of the channel due to the complicated rate region formulae. Numerical examples demonstrated that both users
 can achieve positive rates simultaneously under the information-theoretic secrecy requirement.%\bibliographystyle{IEEEtran}

\bibliographystyle{IEEEtran}

\renewcommand{\baselinestretch}{0.8}
\bibliography{IEEEabrv,SecrecyPs2}

\begin{thebibliography}{10}
\providecommand{\url}[1]{#1}
\csname url@rmstyle\endcsname
\providecommand{\newblock}{\relax}
\providecommand{\bibinfo}[2]{#2}
\providecommand\BIBentrySTDinterwordspacing{\spaceskip=0pt\relax}
\providecommand\BIBentryALTinterwordstretchfactor{4}
\providecommand\BIBentryALTinterwordspacing{\spaceskip=\fontdimen2\font plus
\BIBentryALTinterwordstretchfactor\fontdimen3\font minus
  \fontdimen4\font\relax}
\providecommand\BIBforeignlanguage[2]{{%
\expandafter\ifx\csname l@#1\endcsname\relax
\typeout{** WARNING: IEEEtran.bst: No hyphenation pattern has been}%
\typeout{** loaded for the language `#1'. Using the pattern for}%
\typeout{** the default language instead.}%
\else
\language=\csname l@#1\endcsname
\fi
#2}}

\bibitem{Liang_same_marginal}
Y.~Liang and H.~V. Poor, ``Multiple access channels with confidential
  messages,'' \emph{{IEEE} Trans. Inform. Theory}, vol.~54, pp. 976--1002, Mar.
  2008.

\bibitem{Secureconnect}
X.~Zhou, R.~K. Ganti, and J.~G. Andews, ``Secure wireless network connectivity
  with multi-antenna transmission,'' vol.~10, no.~2, pp. 425--430, Feb. 2011.

\bibitem{Wyner_wiretap}
A.~D. Wyner, ``The wiretap channel,'' \emph{Bell Syst. Tech. J.}, vol.~54, pp.
  1355--1387, 1975.

\bibitem{csiszar1978broadcast}
I.~Csisz{\'a}r and J.~Korner, ``{Broadcast channels with confidential
  messages},'' \emph{{IEEE} Trans. Inform. Theory}, vol.~24, no.~3, pp.
  339--348, 1978.

\bibitem{Shafiee_secrecy_2_2_1_J}
S.~Shafiee and S.~Ulukus, ``Towards the secrecy capacity of the {G}aussian
  {MIMO} wire-tap channel: the 2-2-1 channel,'' \emph{{IEEE} Trans. Inform.
  Theory}, vol.~55, no.~9, pp. 4033--4039, Sept. 2009.

\bibitem{Khisti_MIMOME}
A.~Khisti and G.~W. Wornell, ``Secure transmission with multiple antennas-{II}:
  The {MIMOME} wiretap channel,'' \emph{{IEEE} Trans. Inform. Theory}, vol.~56,
  no.~11, pp. 5515--5532, Nov 2010.

\bibitem{Oggier_MIMOME}
F.~Oggier and B.~Hassibi, ``The secrecy capacity of the {MIMO} wiretap
  channel,'' \emph{{IEEE} Trans. Inform. Theory}, vol.~57, no.~8, Aug. 2011.

\bibitem{Liu_MIMO_wiretap}
T.~Liu and S.~S. (Shitz), ``A note on the secrecy capacity of the
  multiple-antenna wiretap channel,'' \emph{{IEEE} Trans. Inform. Theory},
  vol.~55, no.~6, pp. 2547--2553, Jun. 2009.

\bibitem{Liang_fading_secrecy}
Y.~Liang, V.~Poor, and S.~S. (Shitz), ``Secure communication over fading
  channels,'' \emph{{IEEE} Trans. Inform. Theory}, vol.~54, no.~6, pp.
  2470--2492, Jun. 2008.

\bibitem{Khisti_MISOME}
A.~Khisti and G.~W. Wornell, ``Secure transmission with multiple antennas-{I}:
  The {MISOME} wiretap channel,'' \emph{{IEEE} Trans. Inform. Theory}, vol.~56,
  no.~7, pp. 3088--3104, July 2010.

\bibitem{Goel_AN}
S.~Goel and R.~Negi, ``Guaranteeing secrecy using artificial noise,''
  \emph{{IEEE} Trans. Wireless Commun.}, vol.~7, no.~6, pp. 2180--2189, June
  2008.

\bibitem{Pulu_ergodic}
J.~Li and A.~Petropulu, ``On ergodic secrecy rate for {G}aussian {MISO} wiretap
  channels,'' \emph{{IEEE} Trans. Wireless Commun.}, vol.~10, no.~4, pp.
  1176--1187, Apr. 2011.

\bibitem{Li_fading_secrecy_j}
Z.~Li, R.~Yates, and W.~Trappe, ``Achieving secret communication for fast
  {R}ayleigh fading channels,'' \emph{{IEEE} Trans. Wireless Commun.}, vol.~9,
  no.~9, pp. 2792 -- 2799, Sep. 2010.

\bibitem{gopala2008secrecy}
P.~Gopala, L.~Lai, and H.~El~Gamal, ``{On the secrecy capacity of fading
  channels},'' \emph{{IEEE} Trans. Inform. Theory}, vol.~54, no.~10, pp.
  4687--4698, Oct. 2008.

\bibitem{Liu_MISO}
R.~Liu and V.~Poor, ``Secrecy capacity region of a multiple-antenna {G}aussian
  broadcast channel with confidential messages,'' vol.~55, no.~3, pp.
  1235--1248, Mar. 2009.

\bibitem{Gelfand_Pinsker}
S.~I. Gelfand and M.~S. Pinsker, ``Coding for channel with random parameters,''
  \emph{Problems of control and information theory}, vol.~9, no.~1, pp. 19--31,
  1980.

\bibitem{caire_channel_with_SI}
G.~Caire and S.~Shamai, ``On the capacity of some channels with channel state
  information,'' vol.~45, no.~6, pp. 2007--2019, Sept. 1999.

\bibitem{SC_TIFS}
S.~C. Lin and P.~H. Lin, ``On ergodic secrecy capacity of multiple input
  wiretap channel with statistical {CSIT},''
  \emph{http://arxiv.org/pdf/1201.2868.pdf}.

\bibitem{Kim_lecture}
A.~E. Gamal and Y.~H. Kim, \emph{Lecture Notes on Network Information
  Theory}.\hskip 1em plus 0.5em minus 0.4em\relax
  http://arxiv.org/abs/1001.3404.

\bibitem{Liu_SISO}
R.~Liu, I.~Maric, P.~Spasojevic, and R.~D. Yates, ``Discrete memoryless
  interference and broadcast channels with channels with confidential messages:
  Secrecy rate regions,'' \emph{{IEEE} Trans. Inform. Theory}, vol.~54, no.~6,
  pp. 2493--2507, June 2008.

\bibitem{CostaDPC}
M.~H.~M. Costa, ``Writing on dirty paper,'' \emph{{IEEE} Trans. Inform.
  Theory}, vol.~29, pp. 439--441, May 1983.

\bibitem{pslin_CR}
P.-H. Lin, S.-C. Lin, C.-P. Lee, and H.-J. Su, ``Cognitive radio with partial
  channel state information at the transmitter,'' \emph{{IEEE} Trans. Wireless
  Commun.}, vol.~9, no.~11, pp. 3402--3413, Nov. 2010.

\bibitem{Petropulu_MIMOME}
J.~Li and A.~Petropulu, ``Transmitter optimization for achieving secrecy
  capacity in {G}aussian {MIMO} wiretap channels,''
  \emph{http://arxiv.org/abs/0909.2622}.

\bibitem{Gursoy_security_low_SNR}
M.~C. Gursoy, ``Secure communication in the low-snr regime,'' \emph{submitted
  to the IEEE Transactions on Communications}, Oct. 2009.

\end{thebibliography}

\end{document}